\documentclass[aps,prl,twocolumn]{revtex4}

\usepackage{natbib}
\usepackage{graphicx}
\usepackage{amssymb, amsmath}

\newcommand{\aap}{Astron.\ Astrophys.}
\newcommand{\mnras}{Mon.\ Not.\ R.\ Astron.\ Soc.}

\newcommand{\apjl}{Astrophys.\ J.\ Lett.}
\newcommand{\apjs}{Astrophys.\ J.\ Suppl.\ Ser.}
\newcommand{\araa}{Ann.\ Rev.\ Astron.\ Astrophys.}

\begin{document}

\title{Instability windows and evolution of rapidly rotating neutron stars}
\author{Mikhail E. Gusakov$^{1,2}$}
\author{Andrey I. Chugunov$^1$}
\author{Elena M. Kantor$^1$}
\affiliation{$^1$ Ioffe Physical Technical Institute,
Polytekhnicheskaya 26, 194021 St.-Petersburg, Russia
\\
$^2$St.-Petersburg State Polytechnical University,
Polytekhnicheskaya 29, 195251 St.-Petersburg, Russia
}

\begin{abstract}
We consider an instability of rapidly rotating neutron
stars in low-mass X-ray binaries (LMXBs) with respect to
excitation of $r$-modes (which are analogous to Earth's
Rossby waves controlled by the Coriolis force). We argue
that finite temperature effects in the superfluid core of a
neutron star lead to a resonance coupling and enhanced
damping (and hence stability) of oscillation modes at
\textit{certain} stellar temperatures.
%%%%%%%%%%
Using a simple phenomenological model we
%
%construct a simple phenomenological model and
%%%%%%%%
demonstrate that
neutron stars with high spin frequency
%%%%%%%?????!!!!!
may
%%%%%%%
spend a substantial
amount of time at these `resonance' temperatures. This
finding allows us to explain puzzling observations of hot
rapidly rotating neutron stars in LMXBs and to predict a
new class of hot, non-accreting, rapidly rotating neutron
stars, some of which may have already been observed and
tentatively identified as quiescent LMXB (qLMXB)
candidates. We also impose a new \textit{theoretical} limit
on the neutron star spin frequency,  explaining the cut-off
spin frequency $\sim730$~Hz, following from the statistical
analysis of accreting millisecond X-ray pulsars. Besides
explaining the observations, our model provides a new tool
to constrain superdense matter properties comparing
measured and theoretically predicted resonance
temperatures.
\end{abstract}


 \maketitle

\textit{Introduction.--} Neutron stars (NSs) are rotating
compact objects. Rotation allows NSs to support the modes
restored by the Coriolis force, the so-called {\it
inertial} oscillation modes, including {\it $r$-modes}
\cite{ak01}. The $r$-modes, neglecting dissipation, are
subject to gravitationally driven instability at {\it any}
NS spin frequency $\nu$ \cite{andersson98}; the most
unstable are $r$-modes with low multipolarities ($m=2,3$).
Correspondent timescales, $\tau_{\rm GR}<0$, can be
estimated analytically \cite{lom98,ak01}. Dissipation
suppresses the instability to some extent; at temperatures
of interest, dissipation timescales, $\tau_{\rm Diss}>0$,
for $r$-modes are given by the shear viscosity.
%\cite{sy08}.
%, which is especially efficient at low temperatures  \cite{sy08}.
As a result,
the star is predicted to be unstable with respect to
$r$-modes within the ``instability window'', that is a
region of
%high
spin frequency  $\nu$ and  redshifted internal stellar
temperature ($T^\infty$), where $1/\tau_{\rm
GR}+1/\tau_{\rm Diss}<0$ \cite{ak01}.  For NSs observed in
this region the $r$-mode amplitude should increase
exponentially. An amplified $r$-mode rapidly heats up the
star (by dissipation) and brakes stellar rotation
transmitting angular momentum to gravitational radiation
\cite{levin99}. Therefore, the star should quickly leave
the instability window, making vanishingly small the
probability to observe it unstable.
%%%%%%%%
However, some of NSs, which are observed in low-mass X-ray
binaries (LMXBs; LMXB is a binary system consisting of an
NS and a low-mass companion star, which fills the Roche
lobe) fall well outside the stability region
\cite{hah11,hdh12}. Even additional dissipation mechanisms
(Ekman layer, bulk viscosity, etc.) hardly explain the
fastest and warmest sources
\cite{hah11,hdh12,Andersson_etal13} without appeal to an
exotic NS composition (e.g., Ref.\ \cite{rj13}), strong
vortex pinning at the base of the NS crust \cite{km03} or
rather strong ($> 10^{11}$~G) radial magnetic field at the
crust-core boundary \cite{km03,mendell01}, which is much
larger than typical surface magnetic field of NSs in LMXBs
$\sim 10^8$~G.
%For
%example, to stabilize 4U 1608-522 one should increase
%dissipation by three orders of magnitude (!), which is
%unrealistic.

Furthermore, it is generally believed that LMXBs are
progenitors of the fastest rotating stars -- millisecond
pulsars \cite{acrs82}. They supposed to be spun up by
accretion, but the $r$-mode instability can limit their
spin frequency \cite{aks99b,aks99} at $\nu\sim 300-400$~Hz,
making an interpretation of the faster pulsars (such as
PSR~J1748-2446ad rotating at 716~Hz  \cite{hrsfkc06}) very
difficult.

\begin{figure}
    \begin{center}
        \leavevmode
        \includegraphics[width=8.5cm]{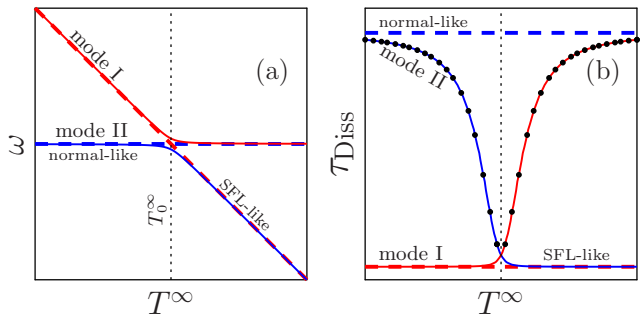}
    \end{center}
    \caption{
    (color online) A scheme of avoided crossing of $r$-mode with $i^s$-mode.
   Oscillation frequency $\omega$ [panel (a)],
   %inverse damping timescale $\tau^{-1}_{\rm Diss}$ [panel (b)],
   and $\tau_{\rm Diss}$ [panel (b)] versus temperature $T^\infty$
   for two oscillation modes (I and II) of a superfluid NS,
   experiencing avoided crossing at $T^\infty=T_0^\infty$.
   Dashes correspond to an approximation of
independent oscillation modes, solid lines show the exact
solution allowing for the interaction of modes I and II.
Vertical dotted lines indicate $T_0^\infty$. Filled circles
in the panel (b) illustrate enhanced damping of
$r$-mode-like oscillations near $T^\infty=T_0^\infty$.
    }
    \label{Fig_scheme}
\end{figure}

To overcome these difficulties, we propose a phenomenological model,
whose cornerstone is superfluidity of neutrons in the NS
core. Theoretical calculations predict that at temperatures
$T^\infty \lesssim 10^8 \div 10^{9}$~K neutrons in the core
are superfluid  \cite{dlz13}, which is confirmed by
observations of cooling isolated NSs
\cite{gkyg05,page04,shternin11,page11}. $r$-modes,
described above, generally are not greatly affected by
superfluidity
 \cite{ly03}.
 % We will call them {\it normal} modes.
They correspond to the comoving oscillations of normal
component (electrons and Bogoliubov excitations of baryons)
and superfluid component (paired neutrons) of the matter;
we will call them {\it normal} modes. However, similarly to
the second sound in superfluids
\cite{khalatnikov89,Sidorenkov_etal13}, an additional class
of inertial modes exists in superfluid stars, -- {\it
superfluid} modes ($i^s$-modes), which are counter-moving
oscillations of normal and superfluid matter components
\cite{ly03,yl03a}. Strictly speaking, $i^s$ and $r$-modes
are clearly distinct only if one sets to zero the so-called
{\it coupling parameter} $s$  \cite{gk11,gkcg13,kg13}.
%When
%$s$ vanishes, equations governing the superfluid and normal
%modes decouple into two independent systems of equations
%\cite{gk11,kg13}.
In this approximation the spectrum and
eigenfunctions of the normal modes coincide with the
corresponding quantities of a nonsuperfluid star and do not
depend on temperature. On the contrary, the
eigenfrequencies of superfluid modes strongly depend on
$T^\infty$ \cite{kg11,cg11}. In reality, the actual
coupling parameter $s$ is although small but finite, and
depends on the equation of state and properties of the
modes \cite{gck13}. This leads to a strong interaction
({\it mixing}) of $r$ and $i^s-$modes when their
frequencies become close to one another. Then the avoided
crossing of modes occurs in the $\omega-T^{\infty}$ plane
[see Fig.\ \ref{Fig_scheme}(a)], similar to avoided
crossings of electron terms in molecules (see, e.g., Ref.\
\cite{ll77}, \S 79). The avoided crossings of oscillation
modes are typical for stellar oscillations
\cite{Asteroseismology,yl03a,gkcg13}; their existence for
$r$-mode and $i^s$-modes is the main ingredient of our
phenomenological model.

\begin{figure}
    \begin{center}
        \leavevmode
        \includegraphics[width=8.7cm]{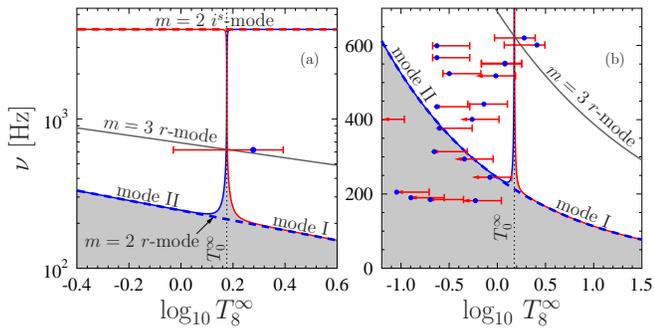}
    \end{center}
    \caption{ (color online)
An example of instability window for superfluid NS. The
star is stable in the grey region, while the white region
is the instability window, splitted up by the `stability
peak' at $T^\infty\approx T_0^\infty$. The solid curves
correspond to instability curves for $m=2$ modes I and II,
respectively, which experience avoided crossing at
$T_0^\infty=1.5 \times 10^8$~K.  The dashed curves
correspond to $m=2$ $r$- and $i^s$-modes  [panel (a) only],
plotted under the assumption that they are completely
decoupled. The grey line is the instability curve for $m=3$
$r$-mode, plotted ignoring the resonance coupling with
superfluid modes. The temperature $T_0^\infty$ is shown by
the vertical dotted line. Panel (b) shows frequencies and
temperatures of the observed sources \cite{Suppl}.
%. The error bars are
%uncertainties due to unconstrained NS envelope composition;
%filled circles give estimation for fiducial crust
%composition.
Only the fastest spinning source 4U~1608-522 is shown in
panel (a).
    }
    \label{f2}
\end{figure}

\begin{figure}
    \begin{center}
        \leavevmode
        \includegraphics[width=7.5cm]{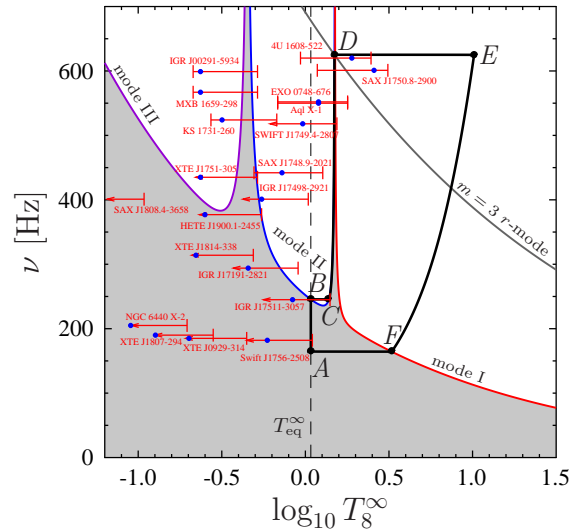}
    \end{center}
    \caption{ (color online)
An example of evolutionary track of an NS in LMXB. The
evolution track $ABCDEFA$ is shown by solid line.
%The star
%is stable in the grey region while the white region is the
%instability window.
The solid curves are instability curves
for $m=2$ modes I, II, and III, respectively; they
experience avoided crossing at $T_0^\infty=4.5 \times
10^7$~K (modes III and II) and $1.5 \times 10^8$~K (modes
II and I).  Other notations are the same as in Fig.\
\ref{f2}.}
%The upper grey line is the instability curve for $m=3$
%$r$-mode, plotted ignoring the resonance coupling with the
%superfluid modes. The temperature $T_\mathrm{eq}^\infty$ is
%shown by the vertical dotted line. Spin frequencies and
%temperatures of the observed sources \cite{Suppl} are
%shown.}
% by filled circles (for
%fiducial envelope composition, see Supplemented Table);
%error bars describe uncertainties arising from
%unconstrained envelope composition.}
    \label{f3}
\end{figure}

\textit{Observational data.--} Spin frequencies $\nu$ and
quiescent effective redshifted surface temperatures
$T^\infty_{\mathrm{eff}}$ are known for 20 neutron stars in
LMXBs
\cite{patruno10,pw12,wsf01,heinke_et_al_10,cackett_et_al_05,dpw12,cackett_et_al_10,muno_et_al_00,cackett_et_al_11,degenaar_et_al_11,cackett_et_al_08,watts12,watts_et_al_08,wijnands_et_al_05,heinke_et_al_09,hjwt07,lowell_et_al_12,rutledge_et_al_99}.
Following Ref.\ \cite{hdh12} we calculate internal
redshifted temperatures $T^\infty$
% (which is constant inside the star)
for each source assuming thermally relaxed crust.
The range of temperatures corresponding to possible envelope
compositions is shown by error bars in Figs.\ \ref{f2},
\ref{f3}; the filled circles correspond to fiducial
composition
(see also supplemented table in \cite{Suppl}).
%; an additional
%source (IGR J17498-2921) and accretion rates are added in
%comparison with\ Ref.\ \cite{hdh12}, along with data
%refinement].
%
Many of the rapidly rotating warm sources fall well outside
the stability region [above the dashed curve in Fig.\
\ref{f2}(b)]
%, plotted under realistic assumptions about the
%properties of superdense matter and
plotted neglecting resonance coupling of $i^s$ and
$r$-modes ($s=0$ approximation).

%%%%%%%%%%%%%%%%%%%%%%%%%%%%5%%%%%%%%%%%%%%%%%%%%%%%%%%%%%%%%%
%%%%%%%%%%%%%%%%%%%%%%%%%%%%%%%%%%%%%%%%%%%%%%%%%%%%%%%%%%%%%

\textit{Mode dissipation in  $s=0$ approximation.--} In
this case, for $r$-modes the dissipation timescale can be
written as $\tau_{\rm Diss}^{r}=\tau_{\rm S}/
\kappa_m\approx \tau_{{\rm S} \, 0} \, \left(T^\infty_8
\right)^2/ \kappa_m$, where $\tau_{\rm S}$ describes
dissipation due to the electron shear viscosity \cite{sy08}
and $\kappa_m$ is a dimensionless coefficient which models
the present uncertainties in the knowledge of shear
viscosity \cite{sy08,bv07,zlz10,sbh13},
%
% (it can vary several-fold depending on a chosen
%EOS, see Fig.\ 1 of
% \cite{sy08}; the neutron shear viscosity can also
%contribute to dissipation  \cite{bv07,zlz10,sbh13}),
and effects of other dissipation mechanisms (such as Ekman
layer dissipation  \cite{ak01,Rieutord01}).
Here $T^\infty_8\equiv T^\infty/(10^8\, {\rm K})$;
$\tau_{{\rm S}\, 0} \approx 2.2 \times 10^5$~s for $m=2$
$r$-mode and $\tau_{{\rm S}\, 0} \approx  1.2 \times
10^5$~s for $m=3$ $r$-mode  \cite{gck13}. To describe
observations, for $m=3$ $r$-mode we take $\kappa_3=5$ as a
fiducial value. For $m=2$ $r$-mode we choose $\kappa_2=1$,
but larger or lower values are also acceptable for our
scenario. For $i^s$-modes damping is very strong
 \cite{yl03a} due to extremely effective \textit{mutual
friction} mechanism, that tends to equalize the velocities
of normal and superfluid components  \cite{als84}. The
corresponding timescale is $\tau_{\rm Diss}^{\rm
sfl}\approx \tau_{{\rm MF} \, 0} (1\,{\rm kHz}/\nu)$,
$\tau_{{\rm MF}\, 0} \approx 4.7$~s  \cite{yl03a}.
Concerning gravitational radiation timescales,
%$r$-modes
%are much better emitters of gravitational waves than
%$i^s$-modes.
for $r$-modes one has $\tau_{\rm GR}^r \approx \tau_{{\rm
GR} \, 0} (\nu/1\,{\rm kHz})^{-2\,m-2}$
 \cite{lom98,ak01} (where $\tau_{{\rm GR} \, 0} \approx
-46.4$~s and $-1250$~s for $m=2$ and $m=3$ $r$-modes,
respectively).
%
%%%%%%!!!!????
For $i^s$-modes gravitational radiation is
suppressed in comparison to $r$-modes
by a factor $c_{\rm GR} \gtrsim 10^4$ \cite{ly03,yl03a}.
For readability of Fig.\ \ref{f2}(a) we take
$\tau_{\rm GR}^{\rm sfl} \approx c_{\rm GR}\tau_{\rm GR}^r$,
with $c_{\rm GR}=100$.
We checked that any $c_{\rm GR} \gtrsim 1$ does not affect our results.
%We checked that our results are insensitive to actual value of $c_{\rm GR}$
%provided that $c_{\rm GR} $\gtrsim 1$.
%%%%%%
%
%For $i^s$-modes gravitational radiation is
%strongly suppressed  \cite{ly03,yl03a} and we take
%$\tau_{\rm GR}^{\rm sfl} \approx c_{\rm GR}\tau_{\rm
%GR}^r$, with $c_{\rm GR}=100$ (note that this parameter can
%be varied by orders of magnitude without affecting our
%results).
%
In $s=0$ approximation, a crossing of modes takes
place in $\omega-T^\infty$ plane (see the dashed lines in
Fig.\ \ref{Fig_scheme}(a)); in that case superfluid and
normal modes would not `feel' each other and the damping
time scales do not have any features associated with
resonant coupling of modes [see the dashed lines in Fig.\
\ref{Fig_scheme}(b)].

\textit{Avoided crossings and dissipation of modes.--} In
reality, coupling of the superfluid and normal modes near
avoided crossings dramatically modifies the dissipation
properties of an oscillating star.  Instead of crossings of
these modes in the $\omega-T^\infty$ plane, one has avoided
crossings: As $T^\infty$ varies, superfluid mode turns into
the normal mode and vice versa (see Fig.\
\ref{Fig_scheme}). For example, the mode II in Fig.\
\ref{Fig_scheme}(a) behaves as normal $r$-mode
(normal-like) at low $T^\infty$ and as a superfluid
$i^s$-mode (SFL-like) at high $T^\infty$. Correspondingly,
its dissipation timescale $\tau_{\rm Diss}$ should smoothly
vary from the high $r$-mode value at $T^\infty<T_0^\infty$
(weak dissipation) to the low value in $i^s$-mode-like
regime at $T^\infty>T_0^\infty$ (strong dissipation). For
the mode I the behavior of $\tau_{\rm Diss}(T^\infty)$ is
opposite [see Fig.\ \ref{Fig_scheme}(b) for
illustration]. The crucial thing about the scenario
proposed in the present note is that $\tau_{\rm Diss}$
should differ substantially before ($T^\infty<T_0^\infty$)
and after ($T^\infty>T_0^\infty$) an avoided crossing,
while the actual form of the function $\tau_{\rm
Diss}(T^\infty)$ at $T^\infty \sim T^\infty_0$ is not
important. Bearing this in mind, in all numerical
calculations we use a simple phenomenological model of the
mode mixing evoked by the perturbation theory of quantum
mechanics.
%!!!!!!!!!!??????????
Within this model the damping
(${\rm X= Diss}$) and gravitational radiation (${\rm X=GR}$) timescales
for the modes I and II
are given, respectively, by
\begin{eqnarray}
 \tau_{\rm X \, I}^{-1} &\approx&
    \sin^2 \theta(x)/\tau_{\rm X}^{r}
  + \cos^2 \theta(x)/\tau_{\rm X}^{\rm sfl},
  \label{1}\\
%\end{equation}
%
%for the mode I and
%and for the mode II
%
%\begin{eqnarray}
\tau_{\rm X \, II}^{-1} &\approx&  \cos^2 \theta(x)/\tau_{\rm
X}^{r}+ \sin^2 \theta(x)/\tau_{\rm X}^{\rm sfl}
\label{2}
\end{eqnarray}
%
%for the mode II
(see Ref.\ \cite{gck13} for more details).
Here
$\theta(x) = [ \pi/2+ \arctan(x)]/2 $
%
%!!!!!?????
and $x \equiv (T^\infty-T_0^\infty)/\Delta T^\infty$; the
parameter $\Delta T^\infty$
%\sim s T_0^\infty$
determines the
%characteristic
width of the avoided crossing.
%To plot Figs.\ \ref{f2} and \ref{f3} we used
%$\Delta T^\infty = 10^{-3} T_0^\infty$.

The qualitative behavior of oscillation modes in superfluid
NSs described above has been confirmed by direct
calculation of oscillation modes in nonrotating NSs
\cite{ga06,kg11,cg11,gkcg13}.
%%%%%%%%%%%%%%%%%%%%%%%
%They demonstrate a large number of resonances between
%normal and superfluid modes which takes place at
%temperatures which are bellow maximal critical temperature
%(superfluid neutrons should exist), but not to low
%(superfluid hydrodynamics should be temperature dependent).
They demonstrate a large number of resonances [well
described by Eqs.\ (1) and (2)] between the normal and
superfluid modes, which occur at $T^\infty \sim
(0.1-1)T_{{\rm c}n}^\infty$, where $T_{{\rm c}n}^{\infty}$
is a typical neutron critical temperature (at $T^\infty
\lesssim 0.1 T_{{\rm c}n}^{\infty}$ we have no avoided
crossings because SFL hydrodynamics is then independent of
temperature).
%%%%%%%%%%%%%%%%%%%%%%%%
Our model is also supported
by the figure 12 of Lee and Yoshida  \cite{ly03}. These
authors employed the zero temperature approximation
($T^\infty=0$) and varied the so-called `entrainment'
parameter $\eta$, that parameterizes interaction between
the superfluid neutrons and superconducting protons. It
follows from the microphysics calculations  \cite{gh05,
gkh09b, gusakov10} that $\eta$ is a function of $T^\infty$.
Hence, its variation is {\it analogous} to a variation of
stellar temperature. Figure 12 of Ref.\  \cite{ly03} shows
the dissipation timescale $\tau_{{\rm MF}}^r$ due to mutual
friction for $m=2$ $r$-mode (or, more accurately, for the
oscillation mode which mimics $r$-mode) as a function of
$\eta$. One can see that $\tau_{{\rm MF}}^r$ sharply
decreases (by few orders of magnitude) at exactly the same
values of $\eta$ at which one observes the resonances
between $r$- and $i^s$-modes in their figure 8, confirming
thus our model. Near the resonances $r$-mode starts to
transform into $i^s$-mode, and hence
 $\tau_{{\rm MF}}^r$
drops down rapidly. Moving away from the avoided crossing
(by decreasing or increasing $\eta$), the solution found by
Lee and Yoshida resembles more and more $m=2$ $r$-mode.
Consequently, $\tau_{{\rm MF}}^r$ grows on both sides of
the resonance, approaching the asymptote value
corresponding to the pure (with no admixture of $i^s$-mode)
$m=2$ $r$-mode [see filled circles in Fig.\
\ref{Fig_scheme}(b)].

%%%%%%%%%%%%%%%%%%%%%%%%%%%%5%%%%%%%%%%%%%%%%%%%%%%%%%%%%%%%%%

\textit{Realistic instability windows.--} The avoided
crossings of modes dramatically modify the instability
window. For instance, assume that modes I and II experience
an avoided crossing at $T^\infty=T_0^\infty=1.5 \times
10^8$~K. Their instability curves, given by the condition
$1/\tau_{\rm GR}+1/\tau_{\rm Diss}=0$,
%for them
are shown
in Fig.\ \ref{f2}(a,b) by solid lines.
%
%%%%%%%%%%!!!!!!?????
To plot the curves we used Eqs.(\ref{1}) and (\ref{2}) and set
$\Delta T^\infty = 10^{-3} T_0^\infty$.
%%%%%%%%%%
%
The panel (b) is a
version of panel (a), but plotted in a different scale. In
addition, Fig.\ \ref{f2}(a,b) shows the instability curves
for: ($i$) $m=3$ $r$-mode; ($ii$) $m=2$ $r$-mode; ($iii$)
superfluid $i^{s}$-mode [panel (a) only]. The latter curves
($i$)--($iii$) are obtained using $s=0$ approximation. As
expected, far from the avoided crossing the solid (modes I
and II) and dashed ($r$- and $i^s$-modes) lines almost
coincide.
%The region of stellar stability (modes I, II, and
%the $m=3$ $r$-mode are simultaneously stable) in Fig.\
%\ref{f2}(a,b) is filled with grey.
The instability window (at least one mode is unstable;
white region) is splitted up by the `stability peak' at
$T^\infty\approx T_0^\infty$ \cite{footnote_curvedPeak}.
This is an inherent feature of $r$- and $i^s$-mode avoided
crossing. In this region the instability curves of modes I
and II continuously change their behavior from
$i^s$-mode-like asymptote to $r$-mode-like one and vice
versa. Therefore, for both modes the instability occurs at
larger frequency, than for pure $m=2$ $r$-mode. The most
unstable mode at $T^\infty=T_0^\infty$ is $m=3$ $r$-mode;
it determines the stability peak height.
In reality, an $r$-mode can experience more than one
avoided crossing with the superfluid modes
\cite{ly03,kg11,cg11,gkcg13}. Unfortunately, the resonant
temperatures $T^\infty_0$ have not yet been calculated
directly; here we treat them as free parameters to be
inferred from observations. In Fig.\ \ref{f3} we
demonstrate the instability windows for two avoided
crossings of $r$-mode with $i^s$-modes -- at $T^\infty=4.5
\times 10^7$~K and $T^\infty=1.5\times10^8$~K.

%%%%%%%%%%%%%%%%%%%%%%%%%%%%5%%%%%%%%%%%%%%%%%%%%%%%%%%%%%%%%%

\textit{NS evolution in LMXB.--} Dramatic modification of
the instability window alters the evolution of an NS in
LMXB.
%
%%%%%%%%%%%!!!!?????
Corresponding equations were derived in Ref.\ \cite{gck13}
and are similar to those obtained in Refs.\ \cite{olcsva98, hl00}
in the absence of $i^s$-modes.
%in the same
%fashion as it was done in the absence of $i^s$-modes \cite{olcsva98, hl00}.
% добавит ссылку на Owen et al. 1998 в literature!!!
They follow from
({\it i}) angular momentum conservation,
({\it ii}) thermal balance of the star, and
({\it iii}) evolution of each mode owing to damping mechanisms
and excitation by gravitational radiation.
The solution to these equations results in the evolution track $ABCDEFA$,
shown by thick solid line in Fig.\ \ref{f3}
(see Ref.\ \cite{gck13} for more details).
%%%%%%%%%%%
%
%Corresponding equations can be derived in the same
%fashion as in the absence of $i^s$-modes \cite{hl00}. They
%follow from ({\it i}) angular momentum conservation, ({\it
%ii}) thermal balance of the star, and ({\it iii}) evolution
%of each mode owing to damping mechanisms and excitation by
%gravitational radiation. The evolution track $ABCDEFA$ is
%shown by thick solid line in Fig.\ \ref{f3}.
%
At the stage $AB$ the star spins up by accretion while
temperature stays constant,
$T^\infty=T_\mathrm{eq}^\infty$, owing to balance between
cooling processes (neutrino emission from the bulk of the
star and thermal electromagnetic radiation from its
surface) and the accretion driven stellar heating
\cite{bbr98}. Being pushed into the instability window by
accretion spin up, the star is rapidly heated up by excited
mode II and reaches the foot of the stability peak in point
$C$.
%%%OLD
%%%%%%%%%%%% \cite{footnote11}.
%%%%%%%%%%%%%
%%%%%%%%%%%%%??????!!!!!!! footnote11
%%%%%%%%%%%%%
%%%%%%%%%%%%Then the star climbs up the stability peak due to accretion
%%%%%%%%%%%%spin up.
%%%%%%%%%%%%%
%%%%%%%%%%%%The stellar temperature is maintained by dissipation of
%%%%%%%%%%%%mode II, which is marginally stable and can be shown to
%%%%%%%%%%%%have the required average amplitude at the stage $CD$.
%%%OLD
%%%%%%%%%%%%%%%%%%
%%%resub3, 13.03.14
%(depending on the parameters of the model
(Note that, for certain parameters of the model a star can execute
the standard `Levin cycle' \cite{levin99} before reaching
the point $C$.)
%
%
% At point C an NS start to move along the boundary of the stability peak;
% the corresponding average amplitude of the mode II
%(for our model it is $\alpha\sim 8\times10^{-7}(400\mbox{\,Hz}/\nu)^4$,
% see equation (46) of Ref.\ \cite{gck13} and discussion around)
% adjusts itself so as to keep
% the star at the boundary.
% so as to force the star move along the boundary of the stability peak
%
At point $C$ the (average) amplitude $\alpha_{\rm eq}$ of the mode II adjusts itself
%so as to force the star move along the boundary of the stability peak.
so as to `stick' the star to the boundary of the stability
peak (as in Refs.\ \cite{ajk02,wagoner02,rb03,no06,gck13}).
Then two alternatives are possible. If gravitational
%andr
wave
%andr
torque corresponding to $\alpha_{\rm eq}$ is larger than
accretion torque, the star will move downwards
\cite{ajk02}; otherwise the star will climb up the peak
\cite{rb03,gck13}. For realistic model parameters adopted
here (see Ref.\ \cite{gck13} for details) the
second possibility is realized and the star reaches the
point $D$ \cite{footnote2}.
%
%\footnote{E.g., for parameters adopted here and in Ref.\ \cite{gck13} it is
% given by $\alpha\sim 8\times10^{-7}(400\mbox{\,Hz}/\nu)^4$ at the stage CD}.
%
%
%Then mode II is excited and have an average amplitude
%required to maintain stellar temperature at the boundary of
%stability peak.
% and can be
%shown to have the required average amplitude
% ($\alpha\sim 8\times10^{-7}(400\mbox{\,Hz}/\nu)^4$)
%at the stage $CD$.
%This amplitude determines gravitation wave braking torque.
%If accretion torque exceeds it, the star climbs up the
%stability peak \cite{gck13}.
%%%resub3, 13.03.14
In point $D$ the star rushes into the instability window,
%\cite{footnote2},
%
%%%%%%%%%%%%%%% footnote %%%%%%%%%%%%%%%%%%%%%%%
%\footnote{In principle, for sufficiently high magnetic
%fields an NS can reach  spin equilibrium \cite{rfs04} (when
%accretion torque balances gravitational and magneto-dipole
%torques) {\it before} approaching the point D. In that case
%it will stay in `equilibrium' point at the peak CD
% until the end of accretion epoch,
% so that the cycle ABCDEFA will never be completed
% (see Ref.\ \cite{gck13} for more details).
%},
%%%%%%%%%%%%%%%%%%%%%%%%%%%%%%%%%%%%%%%%%%%%%%%%
%
where the instability of modes I and $m=3$ $r$-mode converts
rotation energy of the star to gravitational waves and
heat, and brings rapidly the star back to the stability
region in point $F$. Then the star cools down to point $A$
and cycle repeats. The NS spends most of the time climbing
up the stability peak (stage $CD$), i.e. in the region
which is thought to be unstable and unreachable in standard
scenario (i.e., neglecting resonance coupling of $r$- and
$i^s$-modes) \cite{levin99}. Thus, it is not surprising in
our model that we see a number of stars (4U 1608-522, SAX
J1750.8-2900, EXO 0748-676, Aql X-1, and SWIFT
J1749.4-2807) at this stage.

Moreover, the low temperature sources IGR J00291-5934, MXB
1659-298, KS 1731-260, and XTE J1751-305 can not have
$T_\mathrm{eq}^\infty$ larger than that corresponding to
their measured temperatures. If it is low enough, the
evolution track goes along the left edge of low-temperature
stability peak, corresponding to the avoided crossing of
the modes II and III (see Fig.\ \ref{f3}), thus explaining
high spin frequencies of these sources. The rest of the
stars lie in the stability region, so they can be explained
as going through stage $AB$ of their evolution cycle.

\textit{Discussion and conclusions.--} Our simple model
naturally explains rapidly rotating NSs in LMXBs
%%%%%%%%%%%%%%%%%
%within minimal assumptions on NS composition, which are the
%same as in minimal observationally consistent NS cooling
%model \cite{gkyg04}.
within the same assumptions as those adopted in `minimal
cooling' scenarios of Refs.\ \cite{page04, gkyg04} (we note
that these scenarios turn out to be very successful in
interpretation of observations of cooling isolated NSs
\cite{gkyg05, page11, shternin11}).
%%%%%%%%%%%%%%%%%
Furthermore, we predict spin frequencies of NSs to be
limited by the height of the stability peak, defined by the
instability of the $m=3$ $r$-mode, but not by $m=2$
$r$-mode, as it is usually assumed \cite{bildsten98,hdh12}.
This result explains the abrupt observational cut-off of
the spin frequency distribution of accreting millisecond
pulsars above $\sim 730$~Hz  \cite{chakrabarty_etal_03,
chakrabarty08}. In addition, we predict that the main
heating mechanism for the stars climbing up the stability
peak (the most rapidly rotating warm NSs) is vibrational
dissipation, but not accretion, as it is usually supposed
\cite{bbr98}.

Moreover, these NSs will remain attached to the stability
peak $CD$ (and hence stay warm) even after accretion will
be stopped due to, e.g., depletion of the low-mass
companion. Then, according to our scenario, they will start
to spin down slowly (during $10^8-10^9$~years) because of
gravitational and magneto-dipole radiation. Such NSs should
be observed as X-ray sources with purely thermal NS
atmosphere spectrum, and hence should have similar
observational properties as quiescent LMXB (qLMXB)
candidates \cite{Heinke_etal_qLMXB03,Guillot_etal11}. This
means that some of X-ray sources known as qLMXB candidates,
which have never been observed in outbursts
\cite{Guillot_etal11}, can in fact be heated not by
accretion (as is the case for qLMXBs), but by vibrational
dissipation. We propose to call these objects `HOFNARs'
(from HOt and Fast Non-Accreting Rotators) or `hot widows'
(in analogy with the name `black widows' denoting
millisecond pulsars with ablating companions). Note that,
HOFNARs should appear not only in our scenario, but in {\it
any} evolutionary scenario, which assumes that hot rapidly
rotating NSs in LMXBs are located in the instability window
in some (quasi)stationary state (including
low-saturation-amplitude scenarios, see, e.g., Refs.\
\cite{ms13, bw13}).

The temperatures of hot rapidly rotating NSs (both
accreting and non-accreting) must coincide with the
resonance temperatures $T^\infty_0$. The resonance
temperatures, we extract from observations of LMXBs ($\sim
4.5\times 10^7$~K -- $1.5\times 10^8$~K), provide a rough
estimate for neutron critical temperature
$T_{\mathrm{c}n}^{\infty}$ around $2\times
10^8\mathrm{\,K}\lesssim T_{\mathrm{c}n}^{\infty}\lesssim
8\times 10^8$\,K (neutrons should be superfluid at high
temperature resonance, but $T_{\mathrm{c}n}^{\infty}$
should not be too high to guarantee temperature dependence
of $i^s$-mode frequency at low temperature resonance), in
agreement with theoretical predictions  \cite{dlz13} and
constraints imposed by `minimal cooling' scenarios
 \cite{gkyg05,page04,shternin11,page11}. Accurate calculations of
resonance temperatures and new observational data from
future space missions such as SRG  \cite{SRG}, NICER
\cite{NICER}, and LOFT \cite{LOFT} will put much more
stringent constraints on the properties of superdense
matter and parameters of superfluidity.

\begin{acknowledgments}
\textit{Acknowledgments.--} We are grateful to A.~D.
Kaminker, O.~Y. Kargaltsev, G.~G. Pavlov, A.~Y. Potekhin,
Y.~A. Shibanov, A.~I. Tsygan, V.~A.~Urpin, D.~G. Yakovlev,
D.~A. Zyuzin for insightful comments and discussions, and
to O.~V. Zakutnyaya for assistance in preparation of the
manuscript. This work was partially supported by RF
president programme (grants MK-857.2012.2,
%%%resub3, 13.03.14
MK-506.2014.2, and NSh-4035.2012.2),
%%%resub3, 13.03.14
%and NSh-4035.2012.2),
by RFBR (grants 11-02-00253-a, 12-02-31270-mol-a,
%%%resub3, 13.03.14
and 14-02-31616-mol-a
%%%resub3, 13.03.14
%),
by the Dynasty Foundation, and by the Ministry of Education
and Science of Russian Federation (Agreement No.\ 8409,
2012).
\end{acknowledgments}

\onecolumngrid
\appendix
\newpage
%\onecolumn

%\begin{minipage}{\textwidth}
%%%%%%%%%%%%%%%%%%%%%%%%%%%%%%%%%%%%%%%%%%%%%%%%%%%%%%%%%%%%%%%%%%%%%%%%%%%%%%
\section{Supplementary information}
%%%%%%%%%%%%%%%%%%%%%%%%%%%%%%%%%%%%%%%%%%%%%%%%%%%%%%%%%%%%%%%%%%%%%%%%%%%%%%
%%%%%%%%%%%%%%%%%%%%%%%%%%%%%%%%%%%%%%%%%%%%%%%%%%%%%%%%%%%%%

%%%%%%%%%%%%%%%%%%%%%%%%%%%%%%%%%%%%%%%%%%%%%%%%%%%%%%%%%%%%%%%%%%%%%%%%%%%%%%
\begin{table*}[h!]
\caption{
 {\bf Observational data and internal temperatures of neutron stars in low-mass X-ray binaries.} %
%%%%%%%%%%%%%
%\footnote{In comparison with Refs.\ \cite{hdh12,ms13} we
%add an additional source (IGR J17498-2921) and accretion
%rates. We also correct misprint for frequencies for MXB
%1659-298 and KS 1731-260. As in \cite{ms13} we treat
%temperatures in Table 2 of Refs.\ \cite{heinke_et_al_09} as
%local, but not redshifted ones as in Ref.\ \cite{hdh12}. We
%also correct misprint in frequency of SAX J1748.9-2021
%(also known as the first LMXB in NGC 6440
%\cite{heinke_et_al_10}) in Table 3 of Ref.\ \cite{ms13}
%(their frequency $205$~Hz correspond to the second LMXB in
%NGC 6440 \cite{heinke_et_al_10}).}.
%%%%%%%%%%%%%%%%%%
The source names are given in the first column. The second
column presents the neutron-star spin frequencies $\nu$
which are taken from Refs.\ \cite{patruno10,pw12}. The
third column summarizes observational data on neutron-star
redshifted effective temperatures $T_{{\rm eff}}^\infty$ in
the quiescent state. The corresponding values are taken
from the papers quoted in the fourth column. In those
papers the thermal component was fitted by the hydrogen
atmosphere models with the fiducial value of the
neutron-star mass $M=1.4 M_{\odot}$. The internal
temperatures calculated under assumption of a fully
accreted envelope $T^{\infty}_{\rm acc}$, partially
accreted envelope with a layer of accreted light elements
down to a column depth of $10^9$~g~cm$^{-2}$ (the same
fiducial value has been adopted in Refs.\ \cite{%bc09,
hdh12}; the corresponding `fiducial' temperature is
$T^\infty_{\rm fid}$), and purely iron envelope
$T^\infty_{\rm Fe}$ are shown in the fifth, sixth, and
seventh column, respectively. Finally, the eighth column
presents (if available) estimates of the {\it averaged}
accretion rates $\dot{M}$ onto neutron stars and the
corresponding references (the ninth column). The averaging
is performed over a long period of time, which includes
both active and quiescent phases. }
\begin{tabular}{l r r r r r r r r}
\hline
   Source\footnote{%\begin{minipage}{\textwidth}
{In comparison to Refs.\ \cite{hdh12,ms13} we add an
additional source (IGR J17498-2921) and accretion rates to
the table. We also correct misprint in the value of
frequency of MXB 1659-298. As in Ref.\ \cite{ms13} we treat
temperatures in Table 2 of Refs.\ \cite{heinke_et_al_09} as
local surface temperatures, but not redshifted ones as in
Ref.\ \cite{hdh12}. We also correct misprint for the source
NGC 6440 in Table 3 of Ref.\ \cite{ms13} (its frequency
$205$~Hz corresponds to NGC 6440 X-2 -- the second LMXB in
NGC 6440 \cite{heinke_et_al_10}, but the temperature is
given for SAX J1748.9-2021 -- another LMXB in NGC 6440
\cite{cackett_et_al_05}).
%We also provide references to the
%original observational papers reporting neutron star
%temperatures in quiescent state.
} }
               &  $\nu$
           $[\mathrm{Hz}]$
   & $\displaystyle \frac{T^\infty_{\mathrm{eff}}}{10^6\, \mathrm K}$
   & Ref.\
   & $\displaystyle \frac{T^\infty_{\mathrm{acc}}}{10^8\, \mathrm K}$
   & $\displaystyle \frac{T^\infty_{\mathrm{fid}}}{10^8\, \mathrm K}$
   & $\displaystyle \frac{T^\infty_{\mathrm{Fe }}}{10^8\, \mathrm K}$
   & $\displaystyle \frac{\dot M}{M_\odot}$ $[\mathrm{yr^{-1}}]$
   & Ref.\ \\
   \hline\hline
   4U 1608-522          &    $620$    &   $1.51\,\ $        & \cite{rutledge_et_al_99} &   $0.93$        & $1.90$        &    $2.47$        & $3.6\times10^{-10}$& \cite{hjwt07}\\
   SAX J1750.8-2900     &    $601$    &   $1.72\,\ $        & \cite{lowell_et_al_12}   &   $1.18$        & $2.57$        &    $3.11$        & $2\times10^{-10}$& \cite{lowell_et_al_12}\\
   IGR J00291-5934      &    $599$    &   $0.63$\footnote{
   {We treat the effective temperature from the table 2
    of Ref.\ \cite{heinke_et_al_09} as a local one to reproduce the thermal luminosity from that
    reference.}
\label{Heinke_Ts_correction}}
                                                        & \cite{heinke_et_al_09}   &   $0.21$        & $0.24$        &    $0.52$        & $2.5\times10^{-12}$& \cite{heinke_et_al_09}\\
   MXB 1659-298         &    $567$\footnote{According to Refs.\ \cite{wsf01,watts_et_al_08,watts12}}
                                      &   $0.63\,\ $       & \cite{cackett_et_al_08}  &   $0.21$        & $0.24$        &    $0.52$        & $1.7\times10^{-10}$& \cite{hjwt07}\\
   EXO 0748-676 \footnote{\mbox{The radius of this source was fixed at 15.6~km in spectral fits of Ref.\ \cite{degenaar_et_al_11}.}}
                        &    $552$    &   $1.26\,\  $       & \cite{degenaar_et_al_11} &   $0.68$        & $1.20$        &    $1.79$        & \\
   Aql X-1              &    $550$    &   $1.26\,\  $       & \cite{cackett_et_al_11}  &   $0.68$        & $1.20$        &    $1.79$        & $4\times10^{-10}$& \cite{hjwt07}\\
   KS 1731-260          &    $524$\footnote{According to Refs.\ \cite{muno_et_al_00,watts_et_al_08,watts12}}
                                      &   $0.73\,\  $       & \cite{cackett_et_al_10}  &   $0.27$        & $0.32$        &    $0.67$        & $<1.5\times10^{-9}$& \cite{hjwt07}\\
   SWIFT J1749.4-2807   &    $518$    &   $<1.16\,\ $       & \cite{dpw12}             &   $0.59$        & $0.96$        &    $1.54$        & \\
   SAX J1748.9-2021     &    $442$    &   $1.04\,\ $       & \cite{cackett_et_al_05}  &   $0.49$        & $0.72$        &    $1.27$         & $1.8\times10^{-10}$& \cite{hjwt07}\\
   XTE J1751-305        &    $435$    &   $<0.63$\footnotemark[1]     & \cite{heinke_et_al_09}
                                                                                   &   $0.21$        & $0.24$        &    $0.52$        & $6\times10^{-12} $& \cite{heinke_et_al_09}\\
   SAX J1808.4-3658     &    $401$    &   $<0.27$\footnotemark[1]& \cite{heinke_et_al_09}
                                                                                   &   $0.05$        & $0.05$        &    $0.11$        & $9\times10^{-12}$& \cite{heinke_et_al_09}\\
   IGR J17498-2921      &    $401$    &   $<0.93\,\ $       & \cite{dpw12}            &   $0.41$        & $0.55$        &    $1.04$        & \\
   HETE J1900.1-2455    &    $377$    &   $<0.65\,\ $       & \cite{hdh12}             &   $0.22$        & $0.25$        &    $0.55$       & \\
   XTE J1814-338        &    $314$    &   $<0.61$\footnotemark[1] & \cite{heinke_et_al_09}                                                                                    &   $0.20$        & $0.22$        &    $0.49$         & $3\times10^{-12}$& \cite{heinke_et_al_09}\\
   IGR J17191-2821      &    $294$    &   $<0.86\,\ $       & \cite{hdh12}             &   $0.36$        & $0.45$        &    $0.90$        \\
   IGR J17511-3057      &    $245$    &   $<1.1 \,\ $      & \cite{hdh12}              &   $0.54$        & $0.84$        &    $1.40$        \\
   NGC 6440  X-2        &    $205$%\footnote{206 according to \cite{pw12}}
                                      &   $<0.37\,\ $       & \cite{hdh12}             &   $0.09$        & $0.09$        &    $0.20$        & $1.3\times10^{-12}$& \cite{heinke_et_al_10}\\
   XTE J1807-294        &    $190$    &   $<0.45$\footnotemark[1]
                                                        & \cite{heinke_et_al_09}   &   $0.12$        & $0.13$        &    $0.28$       & $<8\times10^{-12}$& \cite{heinke_et_al_09}\\
   XTE J0929-314        &    $185$    &   $<0.58\,\ $       & \cite{wijnands_et_al_05} &   $0.19$        & $0.20$        &    $0.45$        & $<2\times10^{-11}$& \cite{heinke_et_al_09}\\
   Swift J1756-2508     &    $182$    &   $<0.96\,\ $       & \cite{hdh12}             &   $0.43$        & $0.59$        &    $1.10$       &\\
             \hline
\end{tabular}

 \label{Tab_LMXB_Observ}
\end{table*}

%\end{minipage}


\begin{thebibliography}{72}
\bibitem{footnote_curvedPeak}
Generally, $T_0^\infty$ can be a function of $\nu$. This
could make the stability peak a little bit curved, but
would not affect our main conclusions.
\end{thebibliography}
\end{document}